\documentclass{iopart}             % Journal Style
\usepackage[T1]{fontenc}
\usepackage{iopams}
\usepackage{graphicx}
\usepackage{url,graphics,epsfig}
\usepackage{amssymb}
\usepackage[breaklinks=true,colorlinks=true,linkcolor=blue,urlcolor=blue,citecolor=blue]{hyperref}
\usepackage{cite}

\begin{document}
\jl{2}
%
%+++++++++++++++++++++++++++++++++++++++++++++++++++++++++++++++++++++++++++++
%
%  Macro definitions
%
%+++++++++++++++++++++++++++++++++++++++++++++++++++++++++++++++++++++++++++++
\def\etal{{\it et al~}}
\def\newblock{\hskip .11em plus .33em minus .07em}
%+++++++++++++++++++++++++++++++++++++++++++++++++++++++++++++++++++++++++++++
%  End of Macro definitions
%
%+++++++++++++++++++++++++++++++++++++++++++++++++++++++++++++++++++++++++++++
%
%+++++++++++++++++++++++++++++++++++++++++++++++++++++++++++++++++++++++++++++
%
% Title of the paper
%
%+++++++++++++++++++++++++++++++++++++++++++++++++++++++++++++++++++++++++++++
%
\setlength{\arraycolsep}{2.5pt}             % use this for journal style

\title[Formation of SiO by radiative association]
{Formation of silicon monoxide by radiative association: the impact of resonances}

\author{Robert C Forrey$^{1}\footnote[1]{Corresponding author, E-mail: rcf6@psu.edu}$,
	 James F Babb$^{2}$, Phillip C Stancil$^{3}$ and 
	Brendan M McLaughlin$^{2,4}\footnote[2]{Corresponding author, E-mail: bmclaughlin899@btinternet.com}$}

\address{$^{1}$Department of Physics, Pennsylvania State University, 
                             Berks Campus, Reading, PA 19610-6009, USA}

\address{$^{2}$Institute for Theoretical Atomic, Molecular, and Optical Physics (ITAMP),
                          Harvard-Smithsonian Center for Astrophysics, MS-14,
                          Cambridge, MA 02138, USA}

\address{$^{3}$Department of Physics and Astronomy and the Center for Simulational Physics,
\\
	                 University of Georgia, Athens, GA 30602-2451, USA}

\address{$^{4}$Centre for Theoretical Atomic, Molecular and Optical Physics (CTAMOP),
                          School of Mathematics and Physics, The David Bates Building, 7 College Park,
                          Queen's University Belfast, Belfast BT7 1NN, UK}

%+++++++++++++++++++++++++++++++++++++++++++++++++++++++++++++++++++++++++++++
%
%              Abstract
%
%+++++++++++++++++++++++++++++++++++++++++++++++++++++++++++++++++++++++++++++

\begin{abstract}
Detailed quantum chemistry calculations within the multireference
configuration interaction approximation with the Davidson correction (MRCI+Q) 
are presented using an aug-cc-pV6Z basis set, for the potential energy 
curves and transition dipole moments between low lying molecular 
states of singlet spin symmetry for the SiO molecule.
The high quality molecular data are used to obtain
radiative association cross sections and rate coefficients 
for collisions between ground state Si and O atoms.
%Quantal methods are used and compared with semiclassical results.
Quantal calculations are compared with semiclassical results.
Using a quantum kinetic theory of radiative association in which 
quasibound levels are assumed to be in local thermodynamic equilibrium,
we find that resonances play an important role in
enhancing the rate coefficients at low temperatures by several orders 
of magnitude from that predicted by standard quantum scattering
formulations.  These new molecular formation rates may have       
important implications for applications in astrophysics. 
\end{abstract}
%
% insert suggested PACS numbers in braces on next line
%
%\submitto{\jpb}
%\maketitle
% Uncomment for Submitted to journal title message
%\submitted

\vspace{1.0cm}
\begin{flushleft}
Short title: 
%Radiative Association in Si and O atom collisions\\
Formation of SiO by radiative association\\
%\vspace{1.0cm}
%\submitto{\jpb: \today}
J. Phys. B: At. Mol. Opt. Phys. : \today
\end{flushleft}

% Comment out if separate title page not required
\maketitle
%
%++++++++++++++++++++++++++++++++++++++++++++++++++++++++++++++++++++++++++++
%
%      Text of paper follows
%
%++++++++++++++++++++++++++++++++++++++++++++++++++++++++++++++++++++++++++++
\section{Introduction}

The formation of molecules through radiative association can be an important 
astrophysical process contributing to chemical evolution in environments that 
are free of dust and hydrogen. In particular, CO and SiO were detected spectrally 
in the ejecta of Type II supernovae (SNe) between about 100 days and 500 days from 
the initial explosions. The first molecular detection was from SN1987A, 
and theoretical studies indicated that the primary mechanism leading to formation 
of CO and SiO is radiative association \cite{Lepp1990,Liu1992,Liu1996}. 
CO and/or SiO has been detected in several subsequent supernovae \cite{Cherch2011}. 
The radiative association of C and O has been investigated in some detail
\cite{Dalgarno1990,Franz2011,Antipov2013}.
%and, except for differences related to the available molecular data, 
%there is good agreement for temperatures above 1000~K.
It is well known that CO and SiO play fundamental roles
in the dust formation process, though SN1987A remains the best-studied 
event due to its relative proximity. The formation of SiO is thought 
to be a key step in the subsequent formation 
of silicates and dust (e.g. \cite{Marassi2015}).

The rate coefficient for the radiative association of Si and O,
\begin{equation}
\rm Si + O \rightarrow SiO + h\nu,
\end{equation}
where $h\nu$ represents the emitted photon,
was calculated by Andreazza, Marinho, and Singh~\cite{Andreazza1995}
using a semiclassical theory with empirical molecular data.
In this paper, we investigate the radiative association 
of ground state Si and O using accurate molecular data and
several theoretical approaches for the calculation of radiative 
association cross sections. A semiclassical approach to calculating the 
rate coefficient  \cite{Bates1951} is appropriate for heavy atoms such 
as Si and O and may provide a suitable estimate of the non-resonant
contribution for temperatures as low as several hundred degrees Kelvin (K). 
It does not, however, without additional treatment, take into account 
resonances that can occur for certain values of the collision energy. 
Standard quantum-mechanical methods for calculating the cross sections 
are also available, which intrinsically include resonance effects. 
These methods generally add radiative broadening in order to exclude 
very narrow resonances from contributing to the rate constant 
\cite{Antipov2013,Bain1972,Bennett2003,Mrugala2003}. 
Recently, a quantum kinetic theory \cite{Forrey2015} showed that the narrow
resonances should not be neglected and predicted 
%that the narrow resonances 
they
would generally lead to large enhancements in molecular formation rates.
Here, we investigate this possibility for Si and O collisions for the
conditions of local thermodynamic equilibrium (LTE).

%
%##########################################################################################
%
%       Theory section of the paper
%
%##########################################################################################
%
%
\section{Molecular Structure}\label{sec:Theory}
In the present study the potential energy curves and transition dipole matrix elements 
of the low lying electronic states are calculated by using a 
state-averaged-multi-configuration-self-consistent-field (SA-MCSCF) approach, 
followed by multi-reference configuration interaction (MRCI) calculations 
together with the Davidson correction (MRCI+Q)  \cite{Helgaker2000}. 
The SA-MCSCF method is used as the reference wavefunction for the MRCI calculations 
in all our work. Low lying singlet electronic states and the transition dipole matrix 
elements connecting these states are calculated and used in subsequent dynamical 
calculations for the radiative association process.  
Potential energy curves (PECs) and transition dipole moments (TDMs) as a function 
of internuclear distance $R$ are computed out to a bond separation of $R=20$ Bohr. 
%At bond distances beyond this we use a multipole expansion to represent 
%the long-range part of the potentials for this molecule.
The basis sets used in our work are the augmented correlation consistent
polarized aug-cc-pV6Z (AV6Z) basis sets \cite{Kendall1992,Dunning1993,Dunning1999}. 
All the PEC and TDM calculations were performed with the quantum chemistry
MOLPRO 2012.1.21  program package \cite{Werner2012}  running on parallel architectures.
	%
	%	Fig 1
	%
\begin{figure}
\vspace*{.5in}
\centerline{\epsfxsize=4in\epsfbox{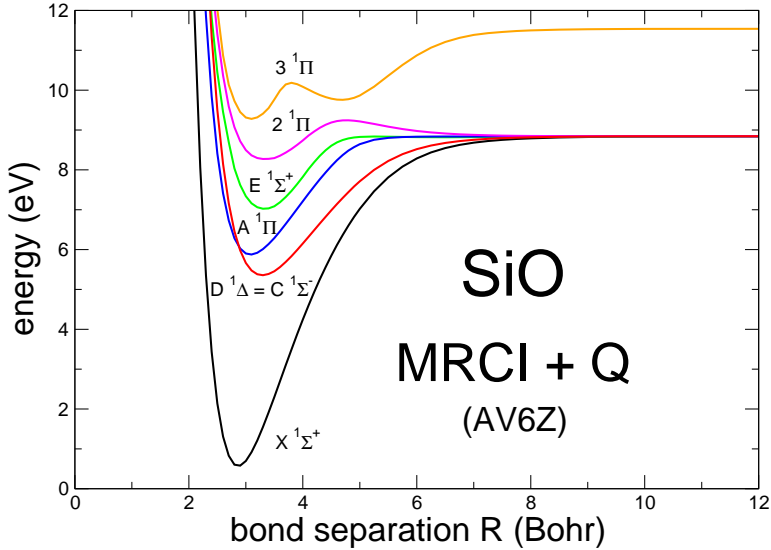}}
\caption{Relative electronic energies (eV) for the  SiO molecule as a 
function of bond separation at the MRCI+Q level of approximation with an AV6Z basis. 
The X$^1\Sigma^+$ ground state along with several of the low lying states,
for the singlet symmetries; C$^1\Sigma^{-}$, E$^1\Sigma^{+}$,  A$^1\Pi$, 2$^1\Pi$ 
and  3$^1\Pi$  are illustrated.}
\label{fig1}
\end{figure}
 	%
        %
        %      Fig 2
 	%
\begin{figure}
\centerline{\epsfxsize=4in\epsfbox{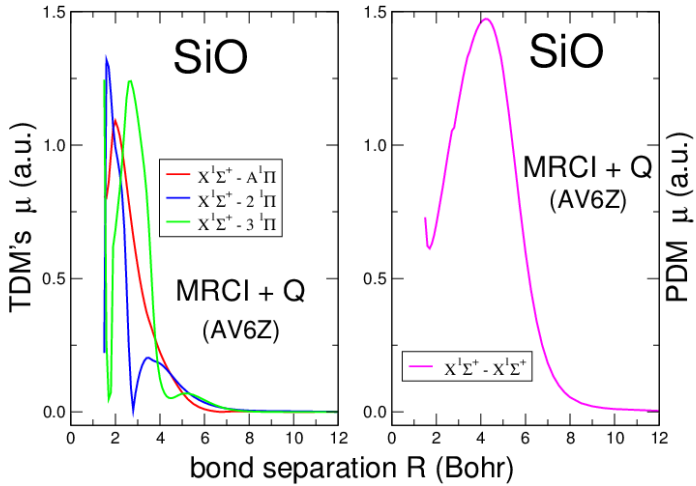}}
\caption{Absolute magnitude of the SiO transition dipole moments (TDMs), 
from the ground state, the left panel is $\mu(R)$ for transitions to the 
low lying singlet $\Pi$ states namely; ${\rm X}^{1}\Sigma^{+} \rightarrow A^{1}\Pi$, 
$2^{1}\Pi$, and $3^{1}\Pi$. The right panel is the absolute magnitude of the 
permanent dipole moment (PDM) for the ${\rm X}^{1}\Sigma^{+}$ ground state of SiO. 
The  MRCI + Q approximation is used with  an AV6Z basis set within the MOLPRO suite 
of codes to determine all the moments as a function of the internuclear distance $R$.}
\label{fig2}
\end{figure}
  	%
        %
%Since for molecules with degenerate symmetry, an Abelian subgroup must
%be used, therefore for a diatomic molecule like SiO with C$_{{\infty}v}$ symmetry, it will be 
The computations were performed in the Abelian subgroup C$_{2v}$
%as the Gaussian orbital based MOLPRO is limited to this highest symmetry. 
with the order of
irreducible representations being ($A_1$, $B_1$, $B_2$,  $A_2$).
We note that the full natural Abelian symmetry
group for a diatomic molecule like SiO is C$_{{\infty}v}$ symmetry.
When symmetry is reduced from C$_{{\infty}v}$ to C$_{2v}$, the correlating relationships
are $\sigma \rightarrow a_1$, $\pi \rightarrow$ ($b_1$, $b_2$) , $\delta \rightarrow$ ($a_1$, $a_2$). 
In order to take account of short-range interactions  we employed the 
non-relativistic state-averaged complete active-space-self-consistent- field (SA-CASSCF)/MRCI method~\cite{Werner1985,Knowles1985} 
available within the MOLPRO~\cite{Werner2012} {\it ab initio} quantum chemistry suite of codes.  
In detail, for the SiO molecule, eight molecular orbitals
(MOs) are put into the active space, including four $a_1$, two $b_1$ and
two $ b_2$ symmetry MOs which correspond to the $3s3p$ shell of Si and
$2s2p$ shell of O atoms. The rest of the electrons in the SiO molecule
are put into the closed-shell orbitals, including four $a_1$, one $b_1$ and
one  $b_2$ symmetry MOs.  The  molecular orbitals for the MRCI procedure 
are obtained from the 
SA-MCSCF method,  where 
the averaging process was carried out on the lowest three $^1\Sigma$ 
($^1A_1$), three  $^1\Pi$ ($^1B_1$), and one $^1\Delta$ ($^1A_2$) 
molecular states of this molecule
(note: the $^1\Delta$ and $^1\Sigma^-$ states are degenerate).
 We use these fourteen MOs ($8a_1$, $3b_1$, $3b_2$, $0a_2$), i.e. (8,3,3,0), to perform 
all the PEC calculations of these electronic states in the MRCI+Q approximation.  Fig \ref{fig1} 
shows the low lying singlet states 
%shows the low lying states of singlet symmetry 
%for the SiO molecule as the molecule dissociates.  
for SiO to their respective dissociation limits.

\newpage

In order to accurately determine the PECs, the point spacing
intervals used here is  0.1 a$_0$ (0.052918 \AA) for each electronic state, 
out to 5 a$_0$ except
near the equilibrium internuclear separation where the spacing is
0.01 a$_0$. Here, a smaller step size is adopted around the equilibrium
position of each electronic state so that the properties of
each PEC can be displayed more clearly. For the present internuclear
distances from 1.5 to 5  a$_0$, the PEC of each electronic
state obtained is smooth and convergent. 
Beyond a bond separation of 5 a$_0$ a spacing of 0.25 a$_0$ 
was used.  It is clearly seen that the two
atoms, Si and O, are completely separated beyond about 10 a$_0$. 
In Fig \ref{fig2} we present the dipole moments as a function of $R$.  
The left panel of  Fig \ref{fig2} illustrates the absolute magnitude of the 
TDMs connecting the low lying singlet $\Pi$
states to the ground state.  The right panel of Fig \ref{fig2} shows 
the absolute magnitude of the permanent dipole moment  (PDM) for 
the X$^1\Sigma^+$ ground state as a function of $R$.  
We note as the molecule dissociates,  both the TDMs
and the PDM are smooth functions of $R$.
The calculated potential energies as a function of $R$ show excellent agreement 
with previous  relativistic-core-potential (RECP) based multi-reference single 
and double CI (MRDCI) calculations \cite{Das2003}, MRCI+Q calculations \cite{Shi2012} 
and  results from the EXOMOL project \cite{Barton2013}.

        %
        %   Table I
        %

\begin{table*}
\centering
\caption{Spectroscopic constants for the X$^1\Sigma^+$ and the A$^1\Pi$ states of the SiO molecule.  
             The equilibrium bond distance $r_e$ and
             the dissociation energies $D_e$ are presented.  
             The present  MRCI + Q results shown are compared with
               previous theoretical work  and  with experiments.}
\label{tab1}
\begin{tabular}{lccccccccc}
\noalign{\vskip 1mm}
\hline  \hline \noalign{\vskip 1mm} 
Basis                           &State && Method                                &&&$r_e$ (\AA)          &&& $D_e$ (eV)            \\
\hline
\noalign{\vskip 2mm}
                                & X $^1\Sigma^+$&&                      &&&                             &&&                             \\
aug-cc-pV6Z                     &&&MRCI+Q$^a$                           &&&1.5153                       &&&8.2748                       \\
aug-cc-pV5Z                     &&&MRCI+Q$^b$                           &&&1.5169                       &&&8.2443                       \\
aug-cc-pV6Z                     &&&MRCI+Q$^b$                           &&&1.5159                       &&&8.2800                       \\
aug-cc-pV6Z                     &&&MRCI+Q/CV+DK$^b$                     &&&1.5114                       &&&8.3281                       \\
CBS                             &&&MRCI+Q/CV+DK +56$^b$         &&&1.5100                       &&&8.3776                       \\
\\
Si (8s/7p/4d/3f/2g/1h)  &&&CCSDT$^c$                            &&&1.5156                       &&&8.2482                       \\
O(7s/6p/4d/3f/2f/1h)    &&&                                             &&&                             &&&                             \\
\\
STO basis                       &&&SCF+CI$^d$                           &&&1.4960                       &&&8.1000                       \\
STO basis                       &&&MCSCF+CI$^d$                 &&&     --                      &&&8.1500                       \\
\\
cc-pVTZ/D-Gauss-Xfit    &&&CASSCF/MRCI$^e$                      &&&1.5237                       &&&                             \\
                                &&&RSPT2$^f$                            &&&1.5328                       &&&                             \\
\\
Si (7s/6p/4d/1f)                &&&RECP + MRDCI$^g$                     &&&1.5210                       &&&7.8700                       \\
O (4s/4p/1d)                    &&&                                             &&&                             &&&                             \\
\\
Experiment                      &&&                                             &&&1.5097$^h$           &&&8.3368$^h$           \\
                                &&&                                             &&&                             &&&8.26 $\pm$ 0.13$^i$  \\
                                &&&                                             &&&                             &&&8.36 $\pm$ 0.09$^j$ \\
                                &&&                                             &&&                             &&&8.18 $\pm$ 0.30$^k$ \\
%                                &&&                                             &&&                             &&&7.93 $\pm$ 0.13$^l$ \\
\\
                                &A$^1\Pi$&&                             &&&                             &&&                             \\
aug-cc-pV6Z                     &&&MRCI+Q$^a$                           &&&1.6315                       &&&2.9693                       \\
aug-cc-pV5Z                     &&&MRCI+Q$^b$                           &&&1.6309                       &&&2.9662                       \\
aug-cc-pV6Z                     &&&MRCI+Q$^b$                           &&&1.6295                       &&&2.9926                       \\
aug-cc-pV6Z                     &&&MRCI+Q/CV+DK$^b$                     &&&1.6249                       &&&3.0146                       \\
CBS                             &&&MRCI+Q/CV+DK+56$^b$          &&&1.6229                       &&&3.0510                       \\
\\
STO basis                       &&&SCF+CI$^d$                           &&&1.4960                       &&&2.44                 \\
\\
cc-pVTZ/D-Gauss-Xfit    &&&CASSCF/MRCI$^e$                      &&&1.7599                       &&&                             \\
                                &&&RSPT2$^f$                            &&&1.8160                       &&&                             \\

\\
Si (7s/6p/4d/1f)                &&&RECP + MRDCI$^g$                     &&&1.6500                       &&&2.5600                       \\
O (4s/4p/1d)                    &&&                                             &&&                             &&&                             \\
\\
Experiment                      &&&                                             &&&1.6206$^h$           &&&3.0259$^h$           \\
                                &&&                                             &&&                             &&&2.87 $\pm$ 0.3$^k$   \\
\noalign{\vskip 1mm}
\hline\hline
\end{tabular}
\begin{flushleft}
$^a$MRCI+Q,  present work, $^b$MRCI+Q,  \cite{Shi2012}, $^c$CCSDT, \cite{Roueff1997},\\
$^d$SCF+CI and MCSCF+CI,\cite{Langhoff1979}, $^e$CASSCF/MRCI \cite{Kork2013} and $^f$RSPT2\cite{Kork2013}.\\
$^g$RECP + MRDCI \cite{Das2003}, $^h$Experiment,\cite{Huber1979}, \\
$^i$Experiment, \cite{Hildenbrand1972}, $^j$Experiment, \cite{Brewer1969}, $^k$Experiment, \cite{Gaydon1968}.\\
\end{flushleft}
\end{table*}

For the radiative association process we are interested in the 
X$^1\Sigma^+$ and the A$^1\Pi$ molecular states of SiO and the 
transition dipole moment $\mu_{X \rightarrow A}(R)$  connecting 
these states as a function of bond length. 
The present MRCI+Q values for the  X-A transition dipole moment 
as a function of bond length are in excellent agreement with the
earlier self-consistent field plus configuration interaction (SCF+CI) 
calculations \cite{Langhoff1979},
a polarization propagator (SPPA) calculation \cite{Elander1976}, 
%recent RECP-MRDCI,  CASSCF/MRCI  
and Rayleigh Schr\"odinger Perturbation Theory 
to second order (RSPT2) \cite{Das2013,Kork2013}.
From our ab initio work we extracted the molecular constants, $r_e$
the equilibrium bond length and the
dissociation energy $D_e$ for these two states of the 
SiO molecule in order to compare with previous experimental and 
theoretical investigations.  Table \ref{tab1} shows our results  and 
compares them to previous experiments and a selection 
of various theoretical methods. 
%From our results presented in Table \ref{tab1} 
Our results are in excellent agreement with previous MRCI+Q 
calculations of similar or slightly higher quality which have been extrapolated 
to the complete basis set limit (CBS).  They are also in excellent 
agreement with the available experiments. 
The  previous MRCI+Q  results that include the additional core-valence (CV)  and relativisitic 
effects incorporated through the Douglas Kroll method (DK) 
only give  marginally better values for these molecular constants. 
We therefore have confidence in the molecular data 
for the PECs and TDMs to be  used in our dynamical studies.

\section{Rate Constant}

%The potential energy curves and transition dipole moments may be used for 
%dynamical studies of radiative association.
The thermal rate constant (in cm$^3$ s$^{-1}$) at a given temperature $T$
to form a molecule by radiative association is given by
\begin{equation}
k_{\Lambda\rightarrow\Lambda^{\prime}} ={\left( \frac{8}{\mu_{r}\pi}\right )}^{1/2}
{\left( \frac{1}{k_B T}\right)}^{3/2} \int_{0}^{\infty} E\; 
\sigma_{\Lambda\rightarrow\Lambda^{\prime}} (E)\; e^{-E/k_B T} dE\ ,
\label{rateconstant}
\end{equation}
where $\Lambda$ and $\Lambda'$ are the initial and final projections of the
electronic orbital angular momentum of the molecule on the internuclear axis,
$\mu_{r}$ is the reduced mass of the Si+O system, $k_B$ is Boltzmann's constant,
 and $E$ is the translational energy. The cross-section 
$\sigma_{\Lambda\rightarrow\Lambda^{\prime}}(E)$ 
may be calculated using semiclassical or quantal methods.
The rate constant is often divided into a sum of two terms
\begin{equation}
k_{\Lambda\rightarrow\Lambda^{\prime}} =
k_{\Lambda\rightarrow\Lambda^{\prime}}^{dir} +
k_{\Lambda\rightarrow\Lambda^{\prime}}^{res}\ ,
\end{equation}
where $k_{\Lambda\rightarrow\Lambda^{\prime}}^{dir}$ 
is the non-resonant direct contribution which may be obtained from
the semiclassical or quantum cross section,
and $k_{\Lambda\rightarrow\Lambda^{\prime}}^{res}$ 
is the resonant contribution which must be obtained
from a quantal method.
We summarize these methods below.

\subsection{Semiclassical Method}

In the semiclassical approximation \cite{Bates1951,Zygel88,Zygel90},
the cross-section for the radiative association process is given by
\begin{equation}
\sigma^{SC}_{\Lambda\rightarrow\Lambda^{\prime} } (E) =  2 \pi P_{\Lambda}
\sqrt{\frac{2\mu_{r}}{E}} \int_0^{+\infty} 
b~ db \int_{R_c}^{\infty} 
\frac{ \Gamma_{\Lambda\rightarrow\Lambda^{\prime}} (R) \ dR}
{\sqrt{1 - V_{\Lambda }(R)/E - b^2/R^2}}\ ,
\label{semiclass}
\end{equation}
where
\begin{equation}
\Gamma_{\Lambda\rightarrow\Lambda^{\prime}} (R) = 2.03 \times 10^{-6} 
\left(\frac{2 - \delta_{0, \Lambda + \Lambda'}}{2-\delta_{0, \Lambda}}\right)
\nu^3(R)\, |\mu (R)|^2 
\end{equation}
is the radiative transition probability in s$^{-1}$ at the given internuclear distance, 
$R_c$ is the classical turning point for the impact parameter $b$, $V_\Lambda(R)$ is the
potential energy in the entrance channel,
$\nu(R)$ is the photon energy in cm$^{-1}$, and $\mu(R)$ is the transition moment in a.u.  
The statistical weight factor $P_{\Lambda}$ is given by
\begin{equation}
P_{\Lambda} = \frac{(2S_{mol}+1)(2-\delta_{0,\Lambda})}
{(2L_{Si} + 1)(2S_{Si} + 1)(2L_O + 1)(2S_O +1)}\ ,
\end{equation}
where $L_{Si}$, $S_{Si}$, $L_O$, $S_O$, are the electronic orbital and spin angular momenta
of the silicon and oxygen atoms, and $S_{mol}$ is the total spin of the molecular electronic state. 
For the $A^1\Pi\rightarrow X^1\Sigma^+$ transition considered in this work, we obtain $P_{\Pi}=2/81$.

\subsection{Standard Quantum Theory}
The quantum mechanical cross-section 
$\sigma^{QM}_{\Lambda\rightarrow\Lambda^{\prime}}(E)$ 
for the radiative association process can be calculated using perturbation theory 
for the radiative coupling
(see Babb and Dalgarno\cite{Babb1995},  Gianturco and Gori \cite{Franco1996} 
or Babb and Kirby \cite{Babb1998}). The result is
\begin{equation}
 \sigma^{QM}_{\Lambda\rightarrow\Lambda^{\prime}} (E)  = \sum_{v^{\prime}j^{\prime}}^{}\sum_{j}^{}
  \frac{64}{3} \frac{\pi^5\hbar^2}{c^3} \frac{\nu^3}{2\mu_{r}E} P_{\Lambda}S_{j j^{\prime}}
   |M_{\Lambda E j, \Lambda^{\prime} v^{\prime} j^{\prime}}|^2,
\label{quantum}
\end{equation}
where the sum is over initial rotational $j$ and final vibrational $v^{\prime}$
and rotational $j^{\prime}$ quantum numbers. 
$S_{j,j^{\prime}}$ are the appropriate line strengths \cite{Cowan1981,Curtis2003} 
or H\"{o}nl-London factors \cite{Watson2008}, and $c$ is the speed of light.  
$M_{\Lambda E j, \Lambda{^\prime} v^{\prime} j^{\prime}}$ is given by the integral
\begin{equation}
{ M_{\Lambda E j, \Lambda{^\prime} v^{\prime} j^{\prime}}}
=\int_{0}^{\infty} F_{\Lambda E j}(R) \mu(R) 
\Phi_{\Lambda^{\prime} v^{\prime} j^{\prime}} (R)dR.
\label{matrix}
\end{equation} 
The wavefunction  $\Phi_{\Lambda^{\prime} v^{\prime} j^{\prime}} (R)$
is a unit-normalized bound state eigenfunction of the final electronic state,
and $F_{\Lambda Ej} (R)$ is an energy-normalized continuum wavefunction
of the initial electronic state.
These wavefunctions may be computed from their
respective Schr\"odinger equations using 
a grid-based numerical approach
%a numerical integration method
\cite{Franz2011,Antipov2013,Babb1995,Gustafsson2012,Antipov2009}.
%However, a finely spaced energy grid would be required to account for 
%the narrow resonances. 
Complicated resonance structures generally
make it difficult to calculate the rate coefficient (\ref{rateconstant})
using numerical integration \cite{Bennett2003,Gustafsson2012}. Furthermore,
the majority of these resonances are sufficiently narrow that the
probability of tunneling is negligible compared to the radiative
decay probability. Consequently, perturbation theory for extremely narrow 
resonances can yield opacities 
%which violate unitarity \cite{Antipov2013,Bennett2003,Gustafsson2012}.
which are larger than unity \cite{Antipov2013,Bennett2003,Gustafsson2012}.
An optical potential approach which includes radiative broadening
may be used to handle the narrow resonances in order to
ensure unitarity \cite{Mrugala2003,Gustafsson2012}.
This approach may also be used to derive a formula for the
resonance contribution
\cite{Antipov2013,Bain1972,Bennett2003,Mrugala2003,Carrington1972}
\begin{equation}
k_{\Lambda\rightarrow\Lambda^{\prime}}^{res}=\sum_{q}\,K_{q}^{eq}\,
\frac{\Gamma_{q}^{tun}\Gamma_{q}^{rad}}
{\Gamma_{q}^{tun}+\Gamma_{q}^{rad}}\ ,
\label{bain}
\end{equation}
where $q\equiv(v_q,j_q)$ designates the vibrational
and rotational quantum numbers for a quasibound state, 
$\Gamma_{q}^{tun}$ and $\Gamma_{q}^{rad}$ are the respective
tunneling and radiative decay widths, 
and $K_q^{eq}$ is the equilibrium constant 
for the quasibound state given by
\begin{equation}
%K_q^{eq}=\frac{g_q\exp(-E_q/k_BT)}{Q_{Si}Q_OQ_T}
K_q^{eq}=\frac{(2j_q+1)P_{\Lambda}\exp(-E_q/k_BT)}{Q_T}\ .
\label{bob2}
\end{equation}
Here $Q_T$ is the translational partition function, 
and $E_q$ is the energy of the quasibound state.
The key step in the derivation of (\ref{bain})
is the replacement of the tunneling width $\Gamma_{q}^{tun}$
in the Lorentzian of a Breit-Wigner resonance
by the total decay width $\Gamma_{q}^{tun}+\Gamma_{q}^{rad}$.
Equation (\ref{bain}) provides an equilibrium population
of quasibound states when the tunneling probability is large 
compared to the radiative decay width.
For extremely long-lived quasibound states such that
$\Gamma_{q}^{tun}<<\Gamma_{q}^{rad}$, however, the 
ratio in equation (\ref{bain}) goes to zero. 
%This effectively eliminates all narrow resonances
%from contributing to the formation rate.
This implies that all narrow resonances
make negligible contributions to the formation rate.
The standard quantum theory result is obtained
when the resonance contribution (\ref{bain}) is 
added to the direct non-resonance contribution 
computed from equation (\ref{quantum}). 
The Sturmian approach \cite{Forrey2013}
described in the next section provides an
alternative method for calculating the standard
quantum theory rate constant if long-lived quasibound
states are eliminated as in equation (\ref{bain}).

%\newpage

\subsection{Quantum Kinetic Theory}
In this section, we follow the approach described by Forrey \cite{Forrey2015} 
and define the radiative association rate coefficient from
the steady-state solution of a self-consistent master
equation.  The formulation uses the bound and unbound 
energy eigenstates of a Sturmian representation to form 
a complete basis set for both the dynamics and kinetics.
All possible transitions are included in the master equation
which allows the extremely long-lived quasibound states to be
populated through bound-unbound and unbound-unbound
interactions.  The result is a phenomenological rate constant
that includes both direct and indirect (inverse predissociation)
processes given by
\begin{equation}
k_{\Lambda\rightarrow\Lambda^{\prime}}
=\sum_{b,u}\,K_u^{eq}\,(1+\delta_u)\,
\Gamma_{u\rightarrow b}^{rad}
\label{bob1}
\end{equation}
where $b\equiv(v_b,j_b)$ and $u\equiv(v_u,j_u)$ designate 
vibrational and rotational quantum numbers for bound and unbound 
states, respectively. $\Gamma_{u\rightarrow b}^{rad}$ is the 
radiative transition probability, and $K_u^{eq}$ is
the equilibrium constant for the unbound state.
%given by
	%
%\begin{equation}
%K_u^{eq}=\frac{g_u\exp(-E_u/k_BT)}{Q_AQ_BQ_T}
%\label{bob2}
%\end{equation}
	%
%where $g_u=2(2j_u+1)$ for approach on the $^1\Pi$ molecular curve,
%$Q_T$ is the translational partition function, and the atomic 
%partition functions $Q_A$ and $Q_B$ are both equal to 9 
%for Si($^3P$) and O($^3P$) states.
The parameter $\delta_u$ is a non-equilibrium concentration defect
which depends on tunneling lifetimes and may be computed as a
function of the matter and radiation temperatures.  For LTE, 
it was shown \cite{Forrey2015} that the concentration defects are 
identically zero. The resulting expression is then equivalent to 
equation (\ref{rateconstant}) using the cross section
\begin{equation}
\sigma_{\Lambda\rightarrow\Lambda^{\prime}}(E)
=\frac{\pi^2\hbar^3}{\mu_rE}P_{\Lambda}
\sum_{b,u}(2j_u+1)\Gamma_{u\rightarrow b}^{rad}\,\delta(E-E_u)
\label{crossx}
\end{equation}
%which is an identity for Sturmian eigenstates
where the delta function is due to the M\"oller operator 
which transforms a free Sturmian eigenstate
%non-interacting discretized continuum state 
into an interacting unbound state with the same energy. 
%As described above, 
For comparison, we note that
standard quantum theory generally uses equations
(\ref{rateconstant}) and (\ref{quantum}) assuming
a Lorentzian line shape
%the Breit-Wigner resonance formula
%\begin{equation}
%2\pi\delta(E-E_u)\approx \frac{\Gamma_u^{tun}}{(E-E_u)^2
%+\left(\Gamma_u^{tun}/2\right)^2}
%\end{equation}
\begin{equation}
|M_{\Lambda E j, \Lambda^{\prime} v^{\prime} j^{\prime}}|^2
\equiv \frac{\Gamma_q^{tun}/2\pi}{(E-E_q)^2
+\left(\Gamma_q^{tun}/2\right)^2}|M_{q\rightarrow b}|^2
%\approx \delta(E-E_q)|M_{q\rightarrow b}|^2
\end{equation}
to obtain an equation which appears similar to equation (\ref{bob1})
for the unbound subspace consisting of only the quasibound states.
Radiative broadening 
(ie. $\Gamma_q^{tun}\rightarrow\Gamma_q^{tun}+\Gamma_q^{rad}$
in the denominator of the Lorentzian) then yields the 
resonance formula (\ref{bain}).

There are two key differences between the quantum kinetic 
result (\ref{bob1}) and the resonance formula (\ref{bain}): 
%(i) the summation includes the non-resonant
(i) equation (\ref{bob1}) includes the non-resonant
background contribution, and (ii) contributions from long-lived 
%quasibound states are not eliminated from the summation. This second
quasibound states are not eliminated from equation (\ref{bob1}). This second
difference is mathematically equivalent to the neglect of radiative 
broadening in the Lorentzian used to derive equation (\ref{bain}).
However, there is no breakdown of perturbation theory 
or violation of unitarity in the result derived from
quantum kinetic theory \cite{Forrey2015}.
It should be noted that LTE in the present context means 
that the matter and radiation temperatures are the same. 
When this is the case, the steady-state solution of the 
Sturmian master equation is a Boltzmann distribution 
for all unbound states, independent of density
and tunneling lifetime. Therefore, all unbound states, 
including extremely long-lived quasibound states,
should be included in the formation rate constant (\ref{bob1}).  
%When this is the case, all quasibound states will be 
%populated by a Boltzmann distribution, independent of
%density and tunneling lifetime, and should be included 
%in the formation rate. 
For comparison, the resonance formula (\ref{bain}), 
which excludes contributions from long-lived quasibound states, 
may be obtained from quantum kinetic theory as an approximate 
steady-state solution of the master equation for a low density gas 
at zero radiation temperature \cite{Forrey2015}.

\section{Results and Discussion}

In this section, we present results from the semiclassical,
standard quantum, and quantum kinetic theories. For the
standard quantum and quantum kinetic results, we employed 
the Sturmian method \cite{Forrey2015,Forrey2013} to compute the
bound and unbound wavefunctions needed to evaluate the
matrix elements (\ref{matrix}).  Quasibound states which 
have negligible tunneling probability compared to their 
radiative decay probability were not included in
%excluded in the resonant contribution (\ref{bain})
%to the standard quantum result.
the standard quantum result in accordance with equation (\ref{bain}).
The long-lived quasibound states, however, were retained 
in the quantum kinetic theory formula (\ref{bob1}) 
which is evaluated in the present work for LTE conditions.
The radiative transition probability includes spontaneous
and stimulated emission and is given by
\begin{equation}
\Gamma_{u\rightarrow b}^{rad}=\frac{A_{u\rightarrow b}}
{1-e^{-(E_u-E_b)/k_BT}}
\label{prob}
\end{equation}
where
\begin{equation}
A_{u\rightarrow b}=\frac{4}{3c^3}
(E_u-E_b)^3\,S_{j_uj_b}\,|M_{u\rightarrow b}|^2
\label{bob4}
\end{equation}
is the Einstein A-coefficient connecting the bound and unbound eigenstates
which were computed by separately diagonalizing the Hamiltonian matrices
for the X$^1\Sigma^+$ and A$^1\Pi$ electronic states in an $L^2$ 
Sturmian representation \cite{Forrey2013} consisting of 500 
Laguerre polynomial basis functions per partial wave.  
As discussed previously \cite{Forrey2013}, it is sufficient
to use unit-normalized positive energy eigenstates directly
because the usual energy normalization exactly cancels the
equivalent quadrature weights which are needed to express
the integration over unbound states as a discrete sum.
This allows the rate constant given in equation (\ref{bob1})
to be easily computed by multiplying equation (\ref{prob}) by 
the equilibrium constant and summing over the bound and unbound states.

\begin{figure}
%\vspace*{.2in}
%\includegraphics[width=\textwidth]{Fig3.pdf}
\centerline{\epsfxsize=4in\epsfbox{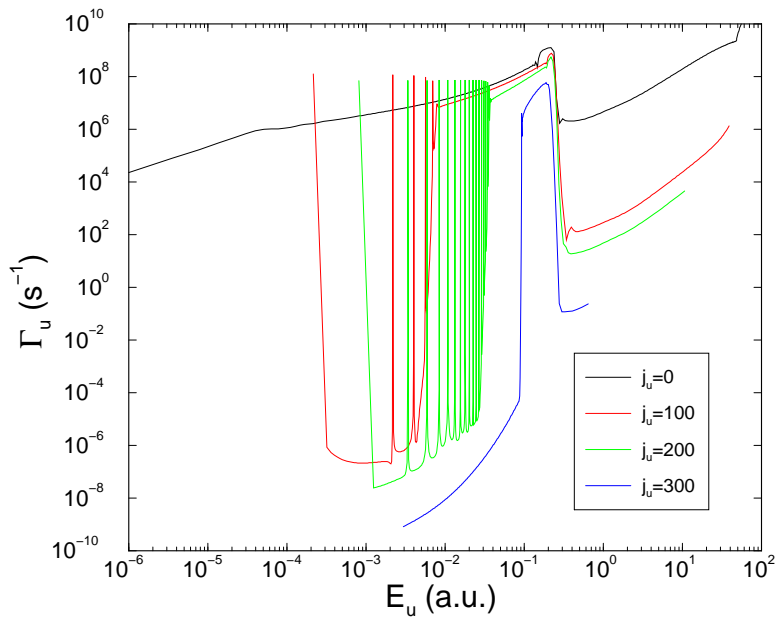}}
\caption{(Colour online) The cumulative radiative width plotted
as a function of the unbound energy  eigenvalues $E_u$ for
$j_u$=0, 100, 200, and 300. The prominant narrow resonant
features are clearly visible in the $j_u$=100 and 200 values.
The cross section can be obtained from the plotted radiative
width via equation (\ref{crossx}).}
\label{fig3}
\end{figure}

Fig \ref{fig3} shows the relative importance of the resonant and
non-resonant contributions for a few partial waves. In the
figure, the cumulative radiative width
\begin{equation}
\Gamma_u=\sum_b\Gamma_{u\rightarrow b}^{rad}
\label{bob3}
\end{equation}
is plotted as a function of the unbound energy $E_u$ for 
zero temperature. 
The figure shows no resonances for $j_u=0$
and an increasing curve due to the increasing energy gap
between the bound and unbound states. The increase stops
abruptly at $E_u\approx 0.2$ a.u. due to the fall-off of
Franck-Condon factors
before further increasing as the energy gap continues to widen.
A similar step-like structure is found near 0.2 a.u. for the 
higher partial waves, however, there are very narrow  
resonances that may be seen at lower energies. The resonances
eventually vanish for large angular momentum (e.g. $j_u=300$).
In our calculations, we found that the last bound state for
the ground state (X$^1\Sigma^+$) potential occurs for 
$j_b=329$, so we included all 500 bound and unbound 
vibrational states for $j_u=0-330$.

        %
        %       Fig 3
        %

\begin{figure}
\vspace*{.1in}
\centerline{\epsfxsize=4in\epsfbox{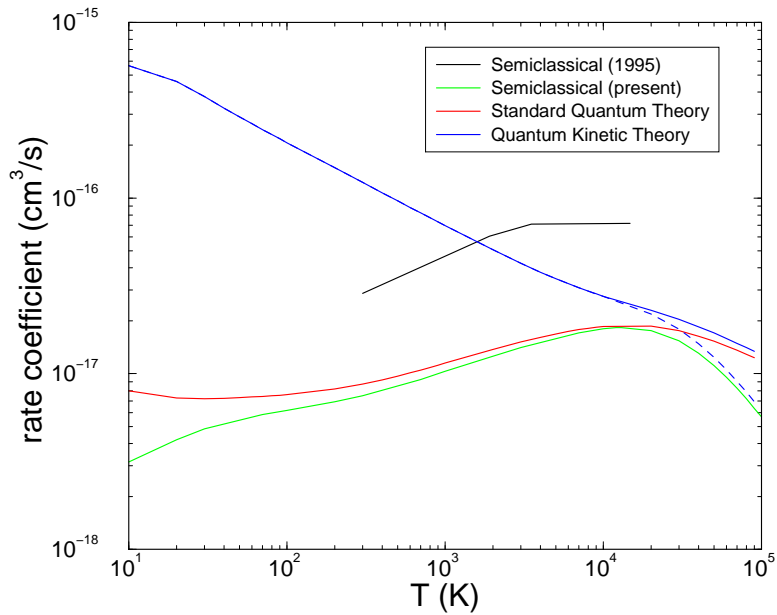}}
\caption{(colour online) Rate coefficient for radiative association
of Si($^3P$) and O($^3P$) via the A$^1\Pi$ molecular curve. 
Present results include semiclassical (green), 
standard quantum (red), and quantum kinetic (blue) curves.
Also shown (black) is the previous semiclassical result 
of Andreazza et al \cite{Andreazza1995}. Note that the 
quantum calculations include stimulated emission.
The semiclassical result and the dashed blue 
quantum kinetic curve include spontaneous emission only.}
\label{fig4}
\end{figure}

Fig \ref{fig4} shows the rate coefficient for radiative association
of Si($^3P$) and O($^3P$) via the A$^1\Pi$ molecular state. The
present results are compared against the previous semiclassical 
result of Andreazza and co-workers \cite{Andreazza1995}.
The standard quantum result was computed from equation (\ref{bob1})
by removing all quasibound states that have negligible tunneling 
probability compared to their radiative decay width. 
This condition is met when the sum
over Franck-Condon factors is greater than 1.99
for the $j\pm 1$ transitions.
Fig \ref{fig4} shows that the standard quantum curve 
is in good agreement with the present semiclassical curve. 
However, both of these curves are significantly smaller than
the previous semiclassical calculation
\footnote{The rate coefficients listed in Table I of \cite{Andreazza1995}
include contributions from E$^1\Sigma^+$, but they are typically
much smaller in magnitude rising to only 41\% of the total rate
at 14,700~K.}.
This difference is
presumably due to the improved molecular data used in the
present work.  When the narrow resonances are included,
the quantum kinetic theory curve shows a difference which
is even more substantial.
The rate coefficient increases with decreasing temperature
and is about 100 times larger than the standard quantum
result which excludes all of the narrow resonances. It should be noted
that the standard quantum curve does include broad resonances, which 
is the cause of the slight upturn in the curve at low temperature.
Stimulated emission is also included in the quantum calculations
\cite{Stancil1997}.
% which accounts for the difference with the 
%semiclassical curve at high temperature. 
When stimulated
association is excluded (e.g. dashed blue curve), the
quantum results merge smoothly with the semiclassical 
calculation at high temperatures.

           %
           %      Fig 4
        %

\section{Conclusions}

We have computed high accuracy SiO molecular structure data
for the purpose of studying radiative association in Si and
O atom collisions. Semiclassical and quantum mechanical
rate coefficients are reported, and the impact of very
narrow resonances is analyzed using a quantum kinetic theory.
The present semiclassical and standard quantum theory results agree 
with each other, however, they are significantly smaller in 
magnitude than a previous semiclassical result. The quantum
kinetic theory includes the extremely narrow resonances
and yields a result which is orders of magnitude larger
than the standard quantum result and has a different
temperature dependence.

The justification for including extremely narrow resonances
comes from the self-consistent master equation
which requires that all quasibound states
are fully populated at LTE. It was further shown 
\cite{Forrey2015,Forrey2016}
that non-LTE conditions may reduce the 
resonant contribution or possibly lead to resonant enhancements 
which are even larger than those reported here. 
This raises an important question about which rate constant
should be used in an astrochemical model. 
The vast majority of astrochemical network models 
are not state-resolved, and the molecules are
assumed to be in their ground state.
Furthermore, the matter and radiation temperatures are
rarely the same in typical low density ISM environments. 
The "standard quantum theory" result corresponds to zero 
radiation temperature at low density and should be used
for modelling such environments. For molecular clouds
which receive light from bright background stars or
are sufficiently dense that quasibound states may be populated
through inelastic collisions, the LTE result would be more 
appropriate.  A more detailed study of the present system 
which includes a full dependence of the resonant contribution 
on both the matter and radiation temperatures is clearly warranted.

In order to compare several collisional theories,
we calculated the rate coefficient for the formation of
%SiO via the process in Eq. (1) along the A${}^1\Pi$ electronic state.
SiO via radiative association along the A${}^1\Pi$ electronic state.
For astrophysical applications, a complete treatment of SiO formation 
from radiative collisions of Si and O must account for approach along
the other electronic states.
%as well as an analysis of inverse predissociation channels.
The increasing formation rate with decreasing temperature
due to the resonances may have important implications 
for astrochemical models of cold molecular clouds.  
%We leave this for a future study.

\ack
RCF acknowledges support from NSF Grant Nos. PHY-1203228
and PHY-1503615. PCS acknowledges support from NASA grant NNX15AI61G. 
BMMcL acknowledges support from NSF through a grant to ITAMP
at the Harvard-Smithsonian Center for Astrophysics under the visitors program,
and Queen's University Belfast for the award of a visiting research fellowship (VRF).
The electronic structure work was carried out at the National Energy Research Scientific
Computing Center in Oakland, CA, USA and at the High Performance Computing Center 
Stuttgart (HLRS) of the University of Stuttgart, Stuttgart, Germany.

%
%+++++++++++++++++++++++++++++++++++++++++++++++++++++++++++++++++++++++++++++
%
%   Reference section now follows
%
%   Delete or change fake bibitem. delete next three
%   lines and directly read in your .bbl file if you use bibtex.
%
%+++++++++++++++++++++++++++++++++++++++++++++++++++++++++++++++++++++++++++++
%
\section*{References}
\bibliographystyle{iopart-num}
\bibliography{sio-ra}

\end{document}